# Machine Learning and statistical classification of CRISPR-Cas12a diagnostic assays

Nathan Khosla[1][#], Jake M. Lesinski[1][#], Marcus Haywood-Alexander[2], Andrew J. deMello[1][*] & Daniel A. Richards[1][*]

[1]Institute for Chemical and Bioengineering, ETH Zurich, Vladimir-Prelog-Weg 1, 8093, Zürich, Switzerland.

[2]Institute of Structural Engineering, ETH Zurich, Stefano-Franciscini-Platz 5, 8049, Zürich, Switzerland.

[#] These authors contributed equally

* daniel.richards@chem.ethz.ch, andrew.demello@chem.ethz.ch,

## Abstract

CRISPR-based diagnostics have gained increasing attention as biosensing tools able to address limitations in contemporary molecular diagnostic tests. To maximise the performance of CRISPR-based assays, much effort has focused on optimizing the chemistry and biology of the biosensing reaction. However, less attention has been paid to improving the techniques used to analyse CRISPR-based diagnostic data. To date, diagnostic decisions typically involve various forms of slope-based classification. Such methods are superior to traditional methods based on assessing absolute signals, but still have limitations. Herein, we establish performance benchmarks (total accuracy, sensitivity, and specificity) using common slope-based methods. We compare the performance of these benchmark methods with three different quadratic empirical distribution function statistical tests, finding significant improvements in diagnostic speed and accuracy when applied to a clinical data set. Two of the three statistical techniques, the Kolmogorov-Smirnov and Anderson-Darling tests, report the lowest time-to-result and highest total test accuracy. Furthermore, we developed a long short-term memory recurrent neural network to classify CRISPR-biosensing data, achieving 100% specificity on our model data set. Finally, we provide guidelines on choosing the classification method and classification method parameters that best suit a diagnostic assay's needs.

# Introduction

Though most known for their role in genetic engineering, Clustered Regularly Interspaced Short Palindromic Repeats–CRISPR-Associated Protein (CRISPR–Cas) systems are increasingly employed as tools for biosensing and in vitro diagnostics (IVDs)[1,2]. Their utility in this regard can be attributed to their ability to specifically target exogenous nucleic acid sequences, such as those associated with invasive pathogens. Since their introduction, CRISPR–Cas-based biosensors have been applied to various diseases[3], including cancers[4], metabolic disorders[5], neurological conditions[6], infectious diseases[7–9], and cardiovascular disease[10]. Success in each area was made possible through iterative optimization of assay chemistries and workflows[11,12]. Beyond assay optimization, Cas enzymes themselves have been engineered to enhance target specificity[1], catalytic activity[13,14], or thermal stability[15,16]. Furthermore, novel chemical- and nanomaterial-based reporters have been synthesized to increase signals[17,18], reduce background[19,20], or simplify assay readout[21,22]. Significantly less effort has been devoted to optimizing how the data emanating from these biosensors is analyzed. This is surprising, given that key assay performance characteristics parameters, such as sensitivity, specificity, and time-to-result are known to be strongly influenced by data analysis methods[19,23]. This is particularly important when employing data processed in real-time, such as that commonly generated from CRISPR–Cas-based biosensors[23]. Furthermore, although machine learning (ML) is well-established within the diagnostic community as a powerful tool for analyzing complex data sets [24,25], the combination of ML and CRISPR–Cas biosensing remains unexplored.

The rapid growth in CRISPR–Cas biosensing was catalyzed by two assays; the DNA Endonuclease-Targeted CRISPR Trans Reporter (DETECTR) assay and the Specific High Sensitivity Enzymatic Reporter UnLOCKing (SHERLOCK) assay[9,26]. Whilst the intricacies of these assays differ, their mechanisms are conceptually similar. In both cases, a Cas protein is programmed to recognize, bind (according to 1:1 kinetics), and cleave a specific target DNA (*cis*-cleavage). After *cis*-cleavage, a conformational change occurs in the Cas protein, resulting in a loss of target specificity and the collateral cleavage of single-stranded DNA (*trans*-cleavage), a process which proceeds according to Michaelis-Menten kinetics (**Figure 1A**). In the presence of a single-stranded DNA reporter containing a fluorophore and a quencher, both assays generate a fluorescence signal proportional to the *trans*-cleavage rate (and determined by the concentration of the initial target nucleic acid)[9,13]. Accordingly, both assays produce time-varying fluorescence signals that can be interpreted in real time. Similarly, to boost assay performance and decrease limits-of-detection, both assays are often coupled with Nucleic Acid Amplification Tests (NAATs).

Slope-based data analysis is the most common method for analyzing fluorescence data obtained from CRISPR–Cas biosensing assays[23,27,28]. For example, Pena *et al.*, developed a real-time method in which the first derivative of time *vs* fluorescence curves is analyzed[28]. In this method, a sample is deemed positive once the slope of the time varying fluorescence signal is greater than a pre-defined threshold (three standard deviations above the maximal negative slope) for three consecutive measurements. Additionally, Fozouni, *et al.* used simple linear regression to determine the slope of fluorescence *vs* time curves, with a sample being deemed positive when its slope is greater than the average negative slope plus two standard deviations (95% confidence interval - CI). Both approaches allow samples to be categorized in real time, while also decreasing the time needed for a positive diagnosis. This has important implications for point-of-care diagnostics, where time-to-result is often a critical performance metric.[29]

Although effective in many situations, slope-driven methods are far from perfect. Numerical differentiations often simply describe a slope at a single point, and thus ignore the broader curve shape. Additionally, single-point methods are highly sensitive to noise, with minor signal fluctuations potentially leading to large variations in the local slope that can significantly impact the ability to precisely categorize samples.[28] Linear regression methods that account for multiple data points circumvent this issue. However, linear regression often misdescribes non-linear data by forcing a linear fit upon the dependent and independent variables. Data from NAAT–CRISPR-Cas assays is the product of several non-linear processes (i.e. exponential amplification, 1:1 binding, Michaelis-Menten kinetics) (**Figure 1A**). Thus, linearity is often only observed during a short portion of the assay. For high-titer samples, the linear phase is typically very short and thus normally described by a very small number of data points. Conversely, for low-titer samples, a linear fit is only possible after collecting many data points, requiring a longer time-to-result. As forcing a linear model onto non-linear data necessarily discards such non-linear effects, more sophisticated methods such as nonparametric statistics or ML are better suited for such analyses (**Figure 1B,C**).

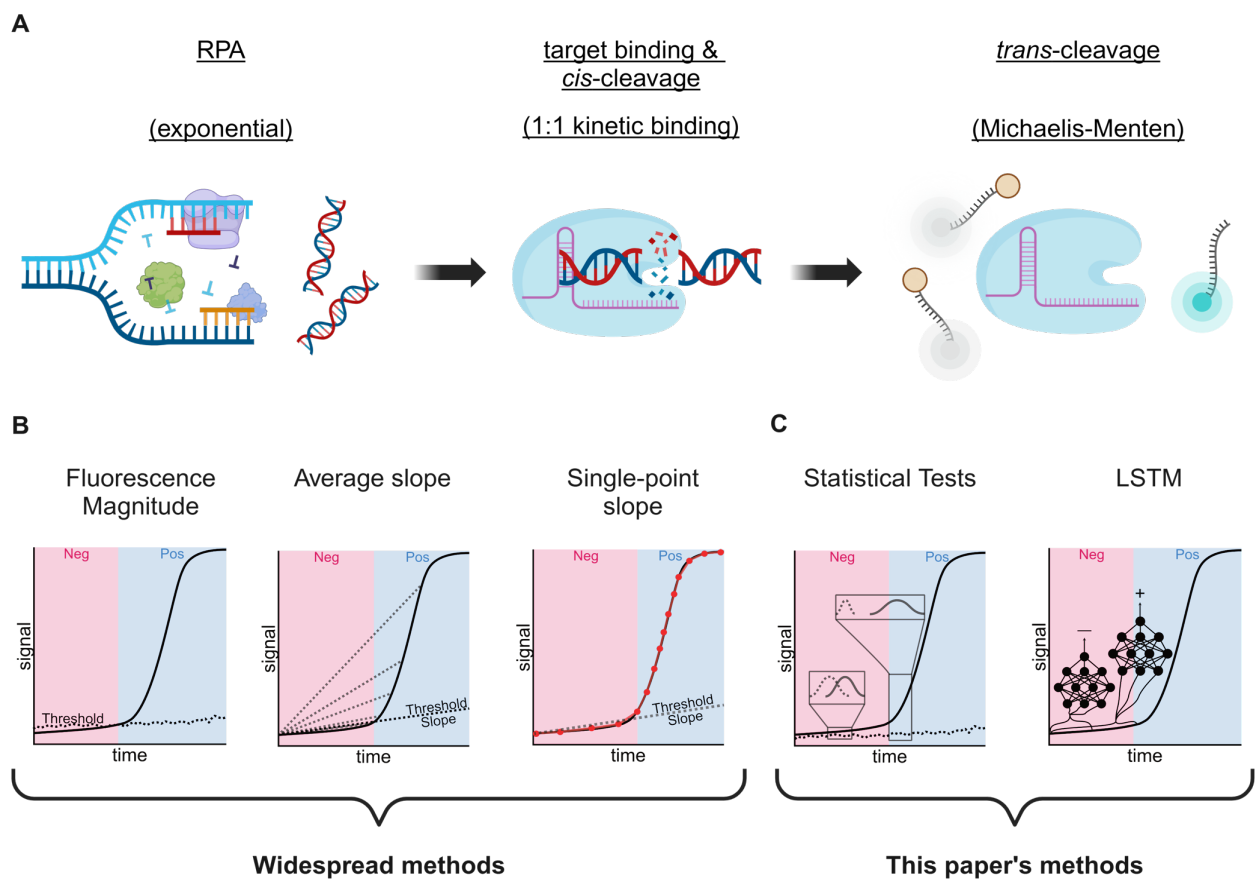

**Figure 1. NAAT–CRISPR–Cas biosensing and the classification algorithms evaluated in this work**. A) Schematic describing the RPA–CRISPR–Cas12a reaction used to generate the data in this study. B) Graphical representations of Fluorescence magnitude, average slope, and single-point

slope methods. C) Graphical representations of non-parametric statistical tests and LSTM neural network presented in this work.

Although rarely used in CRISPR IVDs, nonparametric tests are routinely used to analyze and categorize real-time diagnostic data from lateral flow immunoassays (LFIAs)[30]. Nonparametric tests make no or minimal assumptions regarding the distribution of the data. Thus, they can be more easily applied to smaller or less well-understood data sets. ML approaches have also gained traction in diagnostics, particularly those that employ recurrent neural networks (RNNs), a network architecture in which recent past events are included in computational decisions[31,32]. This capacity for "memory" enables classification and prediction from data trends rather than single data points. Such approaches have been effectively used in real-time NAATs, such as quantitative Polymerase Chain Reaction (qPCR) and quantitative loop-mediated isothermal amplification (qLAMP)[33,34], and LFIAs[35,36], but have yet to be integrated into CRISPR–Cas biosensors[37,38].

Herein, we use non-parametric statistical tests and a Long Short-Term Memory (LSTM) network to analyze time-varying fluorescence data from a CRISPR–Cas biosensing assay and show how these tools can be used to improve both the speed and accuracy of clinical sample classification. Specifically, we employ methods based on a family of non-parametric distribution goodness-of-fit statistics including the Kolmogorov-Smirnov (K-S) test, Anderson-Darling (A-D) test, and Cramér-von Mises (C-vM) test, as well as an LSTM-RNN and apply these methods to data sets obtained from clinical HPV-16 vaginal swab samples. We show that non-parametric statistical methods and the LSTM-RNN consistently outperform traditional analytical methods based on metrics such as sensitivity, specificity, and time-to-result (TTR). This work highlights the currently unexplored potential of integrating statistical analyses and ML into CRISPR–Cas biosensing systems.

## Materials and Methods

**Fluorescence Magnitude Test**

At each time point, analysis was performed without knowledge of any subsequent data. First, a time-dependent negative threshold was determined by taking the mean and standard deviation of the fluorescence data of all 24 negative trials at each time point. The cutoff was determined as the average of the negatives plus three standard deviations at a given time point. A sample was deemed positive when three consecutive fluorescence measurements exceeded the calculated threshold (a run length of 3).

**Average Slope Test**

At each time point, analysis was performed without knowledge of any subsequent data. First, a time-dependent negative average slope threshold was determined for each time, $t_f$, by performing a linear regression from $t=0$ to $t=t_f$ on all the negative samples using the `stats.linregress()` function of the `scipy.stats` Python package. The threshold at each time was then calculated as the returned predicted slope plus three standard deviations of the predicted slope. For sample analysis, the same linear regression process was performed at each time point, and this value compared to the threshold value. A sample was deemed positive when three consecutive slope measurements exceeded the calculated thresholds, as described by Fouzoni et al.[23].

**Single-Point Slope Test**

At each time point, analysis was performed without knowledge of any subsequent data. The maximum negative slope was determined by first finding the single point slope (change in fluorescence divided by change in time) for each measurement point in each of the 24 negative trials on the interval $t_0=0$ to $t_f=90$ minutes and then taking the maximum observed value. The threshold was then calculated as this maximum plus three standard deviations of the set of all observed negative slopes. To analyze a sample, the single point slope (change in fluorescence dividing by change in time) was calculated at each time, and this value compared to the threshold value. A sample was deemed positive when three consecutive slope measurements (a run length of 3) exceeded the calculated thresholds, as described by Pena *et al.*[28]

**Kolmogorov-Smirnov Test**

At each time point, analysis was performed without knowledge of any subsequent data. This test compares two sample populations and returns a statistic indicating the likelihood that the two populations are from the same probability distribution. We constructed two populations; one from the sample to be analysed as well as one from the set of all negative sample readings. In both cases, the populations included only datapoints from the current time and two previous times (a sliding analysis window of length 3). In the case of the test to be analyzed, this resulting population only contained data from that test, whereas the negative population contained datapoints within the sliding window for all 24 known negative tests. The two populations were then compared using a 2-sample, 2-sided Kolmogorov-Smirnov test ($\alpha=0.003$) for goodness of fit from the `ks_2samp()` function of the `scipy.stats` Python package[39,40]. Samples were identified as being positive after $p < \alpha$ once (i.e. requiring a run length of 1). This indicated that the sample was from a different distribution than the reference negative set, with associated probability of a false positive (type I error) equal to $\alpha$. Cut-off values for run length, window length and $\alpha$ threshold were optimized to maximize total accuracy (**Table S2**, **Figures S2-4**).

**Anderson-Darling Test**

At each time point, analysis was performed without knowledge of any subsequent data. As above, two populations were constructed from the sample to be analysed as well as the set of all negative sample readings. The population to be analysed was determined for each time-point using a sliding window of length three. The same was then done for a combination of all negative tests. These two populations were then compared using a k-sample Anderson-Darling[41] test for goodness of fit from the `anderson_ksamp()` function of the `scipy.stats` Python package with α=0.0015. The two populations were deemed to be from different distributions, indicating a positive sample, if $p < α$ once (a run length of 1). Cut-off values for run length, window length and α threshold were optimized to maximize total accuracy (**Table S2**, **Figures S5-7**).

**Cramér-von Mises Test**

At each time point, analysis was performed without knowledge of any subsequent data. As described above, two populations were constructed from the sample to be analysed as well as the set of all negative sample readings. The population to be analysed was determined for each time-point using a sliding window of length six. The same was then done for a combination of all negative tests. These two populations were then compared using a 2-sample, Cramér-von Mises[42] test for goodness of fit from the `cramervonmises_2samp()` function of the `scipy.stats` Python package with α=0.0001. The two populations were deemed to be from different distributions, indicating a positive sample, if $p < α$ once (a run length of 1). Cut-off values for run length, window length and α threshold were optimized to maximize total accuracy (**Table S2**, **Figures S8-10**).

**Long Short-Term Memory Analysis**

An LSTM network is a specialised type of recurrent neural network designed to capture and retain information from sequential data. Here, the LSTM-RNN was applied independently to the time-series fluorescence data from each trial. In training, the objective function in the form of binary cross-entropy loss, $L$, is minimised, for $N$ sequences, i.e.

$$L = -\frac{1}{N}\sum_{i=1}^{N}[y_i \log(\hat{y}_i) + (1 - y_i) \log(1 - \hat{y}_i)]$$

Where $y_i$ is the true label of series $i$, $\hat{y}_i$ is the predicted probability of series $i$. To properly capture the variability in the training stages (and thus in the final model), and to accurately judge the model architecture independent of entropic effects arising from randomness, we trained 10 different models, each with the same architecture but varying the training and testing set selection (as determined by the random number seed used to instantiate the network). To obtain representative values, data were averaged across all seeds. For each seed, data were split into training and validation sets, representing 70% and 30% of the full data, respectively. In this work, the LSTM-RNN was built and trained using PyTorch[43] with a dropout rate of 0.2, 2 layers, a hidden size of 64, and a final sigmoid layer. The predicted label was set as positive (1) when the output of the final layer of the LSTM exceeded the chosen threshold of 0.95 (**Figure S11**).

**Acquisition of the model data set**

For all analyses we used a previously published data set from Lesinski *et al.*[44]. In this work the authors analyzed 16 (8 positive and 8 negative, with triplicate assays of each sample) vaginal swabs for HPV-16. Specifically, we used the RPA-CRISPR Cas12a one-pot assay data. This dataset included the RPA-CRISPR Cas12a raw fluorescence values and Anyplex qPCR values for the clinicals samples.

## Results and Discussion

**Slope- and magnitude-driven classification**

To establish appropriate benchmarks for binary classification in our CRISPR-Cas assay, we first evaluated the previously described slope-based methods reported by Pena *et al.* and Fouzoni *et al.* on the model data set[23,28]. We also included a fluorescence magnitude analysis for comparison, as this is arguably the simplest method for classifying positives and negatives in analytical assays[9,23]. We applied these widespread methods to previously published data obtained from an RPA–CRISPR-Cas assay for detecting HPV-16[44]. The data set included time-varying fluorescence data from 24 positive and 24 negative assays, obtained using clinical vaginal swab samples, as confirmed by Allplex qPCR[44] (**Table S1**). From these analyses, we calculated key performance characteristics, namely sensitivity, specificity, and total accuracy (percentage of samples correctly indicated). These metrics are quantified in **Figure 2**.

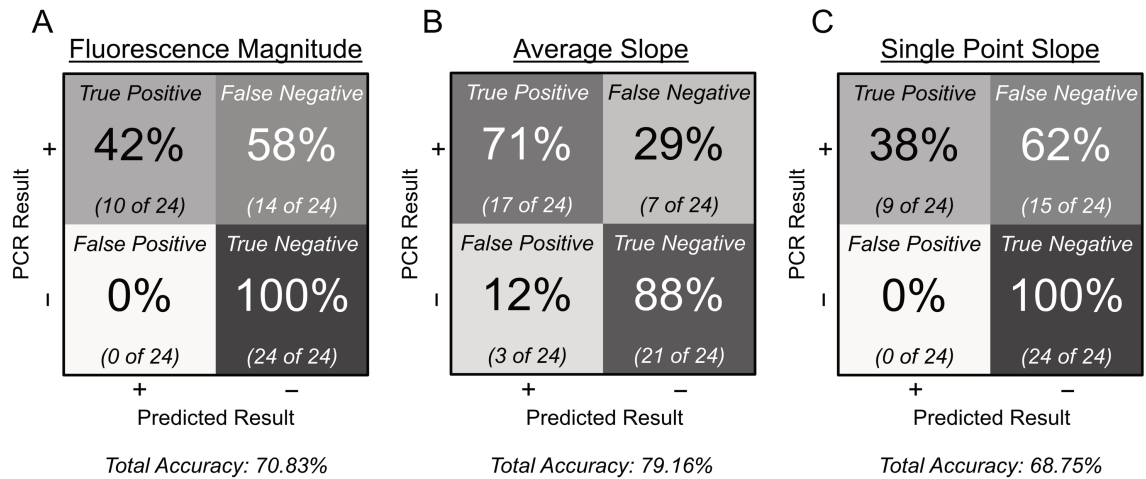

**Figure 2. Performance of widespread slope-based algorithms**. A) Confusion matrix for the fluorescence magnitude method, analysed using a 99.7% CI. B) Confusion matrix for the average method, analysed using a 99.7% CI. C) Confusion matrix for the single point slope method, analysed using a 99.7% CI.

These results highlight several significant differences between the methods. The sensitivity of both the fluorescence magnitude and single-point slope approaches was poor, with respective true positive rates of 42% and 38%. Conversely, both tests displayed perfect specificity, yielding no false positives (100% true negatives). The average slope method yielded a far greater sensitivity of 71%, but was less specific, with a true negative rate of 88%. The average slope algorithm demonstrated a better total accuracy. These data suggests that the fluorescence magnitude and maximal slope algorithms are relatively strict, prioritising greater specificity at the expense of lower sensitivity.

**Non-parametric statistical classification**

After establishing the performance of the fluorescence magnitude and slope-based analysis methods, we next evaluated the sensitivity, specificity, and total accuracy of nonparametric classification methods. We chose nonparametric methods since data from CRISPR–Cas assays cannot be assumed to originate from a known distribution. The K-S test determines the probability of two data sets being sampled from the same distribution by comparing their empirical cumulative distribution functions (ECDFs) (**Figure S1**). The K-S test is a quadratic empirical distribution function (EDF) test, a family of statistical tests that also includes the C-vM test and the A-D test [39]. Though conceptually similar, these tests diverge in how they compare the two distributions. The K-S test compares the maximum difference between the two distribution functions, making it less sensitive to deviations at the tails of the distributions. Conversely, the A-D test uses a weighted sum of the differences between the distribution functions, with more weight given to differences in the data at the extremes of the distributions. The C-vM test compares the squared differences between the distribution functions, integrated over the entire data range. Accordingly, it is sensitive to deviations across the entire distribution[45]. Given these differences, we decided to compare the performance of the three methods on our model data set (**Figure 3**). In each case, empirical distribution functions were constructed from multiple consecutive measurements (window length). To improve robustness, samples were only classified after several consecutive distributions reached significance. We assessed multiple different confidence intervals, window lengths, and run lengths (**Figures S2-S10**), choosing the parameters that provide the greatest total accuracy.

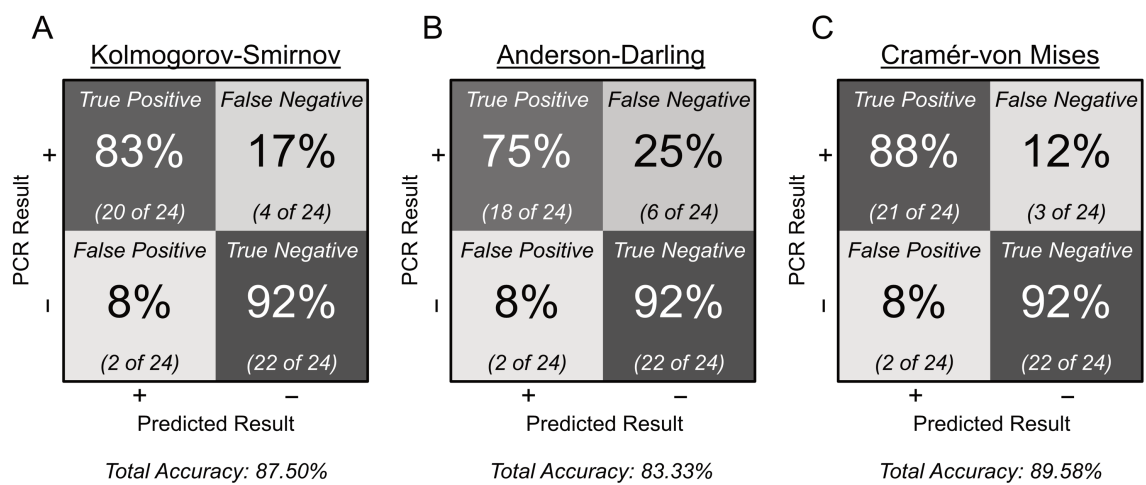

**Figure 3. Performance of quadratic EDF classification methods**. A) Confusion matrix of the K-S test. Confidence interval = 99.7%, window length = 3, run length = 1. B) Confusion matrix of the A-D test. Confidence interval = 99.85%, window length = 3, run length = 1. C) Confusion matrix of the C-vM test. Confidence interval = 99.99%, window length = 6, run length = 1.

As seen in **Figure 3**, all three quadratic EDF classification methods were more sensitive than the slope-based methods, with true positive rates of 83, 75, and 88% for K-S, A-D, and C-vM, respectively. However, this comes at the expense of specificity, as evidenced by the lower true negative rate of 92% observed for all three methods. In terms of total accuracy, the quadratic EDF statistical tests significantly outperform both slope-based and fluorescence magnitude methods, with accuracies of 87.5, 83.33 and 89.58%.

**LSTM classification**

With the growth in accessible computational power, ML methods have proven valuable in classification tasks[46,47]. When considering time-varying data, recurrent neural networks are particularly relevant[48,49]. RNNs are a categorization tool where previous data information can be included in current decision-making processes. RNNs can handle such data due to their capacity for memory. Long-Short-Term-Memory RNNs employ three-way gating mechanisms (input, forget and output gates) to control the flow of information and capture long and short-term trends when making predictions. This is achieved using memory cells and hidden states, with one LSTM cell being used as a node in the classic neural network configuration[50]. Data at each sequential point are passed into the LSTM, along with the cell and hidden states of the previous sequential point. When an LSTM is used for binary classification, the last layer comprises a sigmoidal activation function. Accordingly, the direct output of the LSTM is a value between 0 and 1, representing the probability of the sequential data being the positive (1) class. A more detailed description of Long Short Term Memory RMMs is given by Staudemeyer and Morris[51]. To test the utility of LSTM-RNNs in classifying positive and negative samples in CRISPR–Cas assays, we established an LSTM architecture, optimized the prediction threshold (**Figure S11**), and applied it to our data set (**Figure 4a**). Prior to training, the data set was randomly split into training and validation sets, the composition of which were dependent on the instantiated random number seed. This ensures a more accurate view of the performance of the LSTM by observing the effect of different training set configurations. We observed different performance levels from different seeds and therefore different topologies of the training domain (**Figure S12**). Accordingly, to avoid random bias, we used ten seeds to perform iterations of model training, validation, and data analysis (**Table S3**). Finally, we averaged the performance metrics of all validation sets from the ten models (**Figure 4a**).

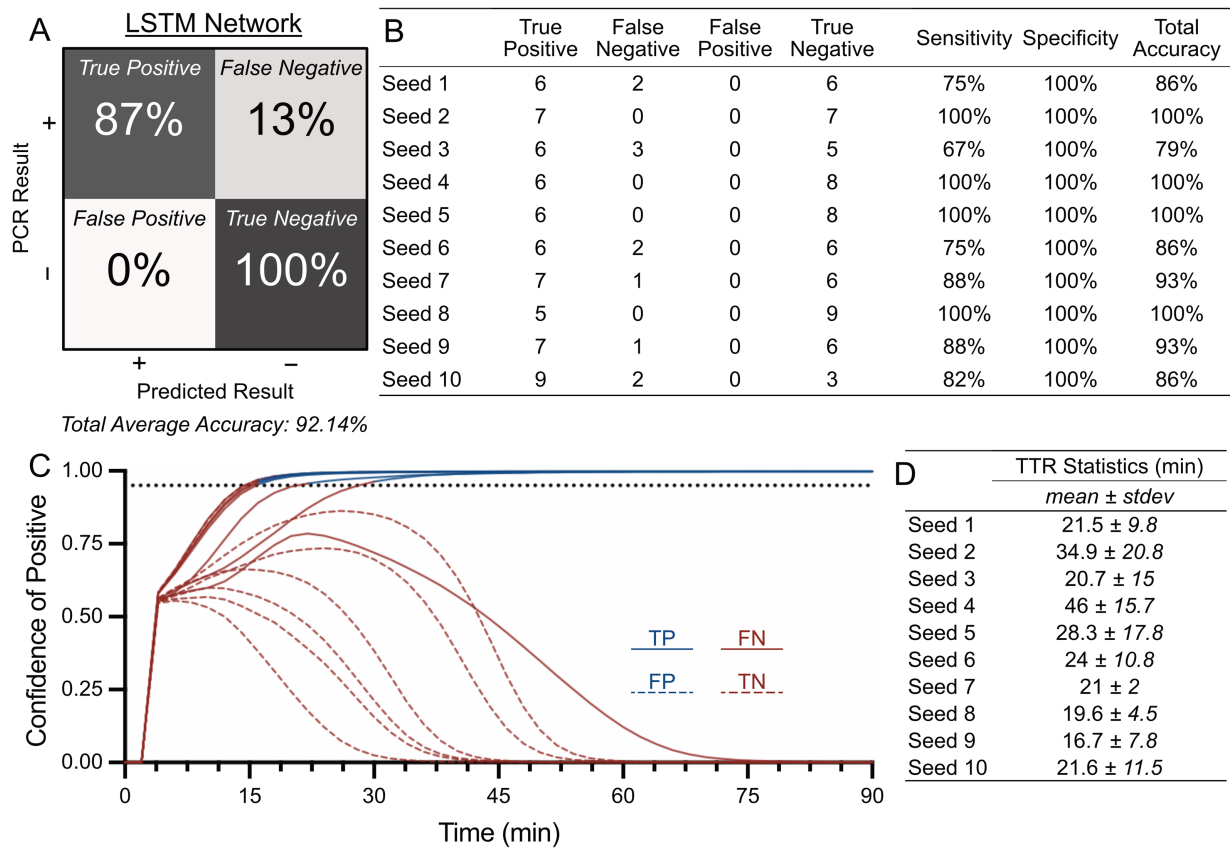

**Figure 4. Performance of LSTM RNNs.** A) Confusion matrix. B) Sensitivity, specificity and accuracy of each LSTM seed. C) LSTM decision/confidence visualization over time for samples in a representative seed. The prediction threshold of 0.95 is indicated with a dotted black horizontal line. Solid lines indicate samples that are actual positives, while dashed lines indicate actual negatives. Samples classified as positive by the LSTM are indicated in blue. This seed reported a single false negative, as shown by the solid red line. The LSTM reported no false positives across all seeds. D) Average time-to-result of each LSTM seed.

We applied the LSTM-RNN to our data set and observed a sensitivity of 87%, specificity of 100% and total accuracy 92.14% (**Figure 4b**). Accordingly, the LSTM outperforms both the slope-based and EDF statistical methods. However, the limited sample size used to train the network does increase the possibility of overfitting the LSTM, distorting performance. This risk was mitigated by training in multiple independent trials with random seeds. An example of the decision process over time for one seed is shown (**Figure 4c**). At each time point, the LSTM analyses data up to that time point and returns a measure of confidence (between 0 and 1) that the data represents a positive test. Confidences above 0.95 were accepted as positive tests, whereas confidence values less than 0.95 were interpreted as a negative test. Further, in assessing performance it is important to account for the variation introduced during the training process by comparing different models trained with different random seeds (**Figure 4D**). In doing so, it becomes apparent that, although the LSTM-RNN may overperform or underperform in each seed, the overall trend is toward improved performance. Application of LSTM-RNNs to other diagnostic systems would naturally require retraining to account for the specific features of each assay, however, our data suggests that this effort would lead to significant improvements in test accuracy.

**Impact of Classification Method on Time-To-Result**

In addition to sensitivity, specificity, and accuracy, TTR is an important measure of assay performance. Obtaining a result within a specific time frame is paramount in many scenarios, such as rapid testing during an infectious disease epidemic/pandemic.[52,53] Accordingly, we investigated how the different classification techniques impact TTR (**Figure 5A,B**). Here, TTR is defined as the first time point at which a given sample matches the positive criterion, according to each individual method.

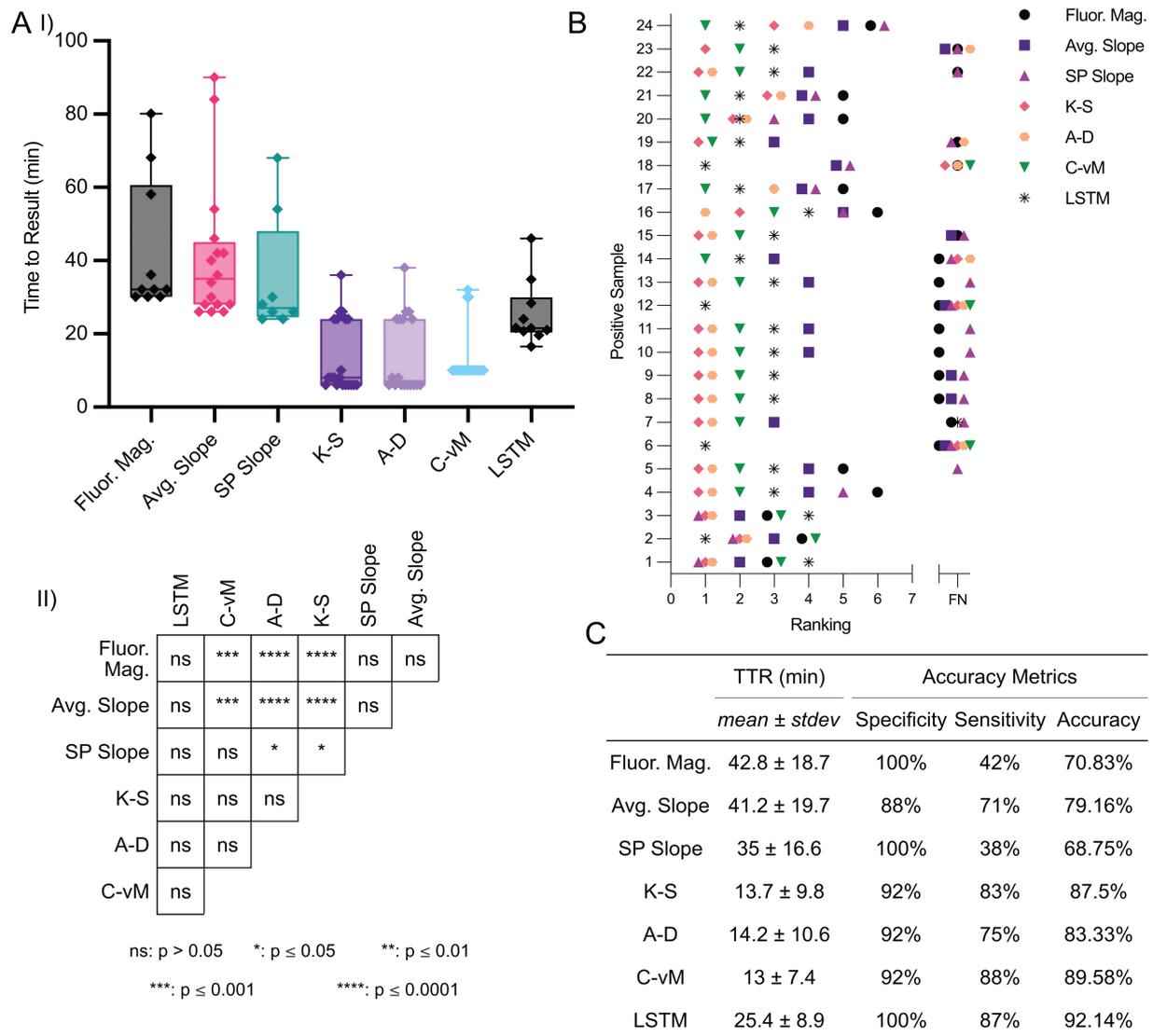

**Figure 5. Comparison of TTR and accuracy metrics across all analysis methods.** A) TTR values for tests analysed using the seven different methods (i) as well as pairwise comparisons for statistically significant differences in the TTR of each test(ii). B) Ranking of each method, from fastest (1) to slowest (7). False negatives are grouped as FN. C) TTR, specificity, sensitivity, and accuracy for each test.

The data in **Figure 5** highlight several important features of each method. The slope-based methods returned the slowest average TTRs, ranging between 35.0 and 42.8 minutes. As shown in **Figure 5C**, the range of TTRs for each method was large; 30-80 minutes, 26-90 minutes, and 24-68 minutes for the fluorescence magnitude, average slope, and single-point slope methods, respectively. The EDF methods returned the fastest average TTRs, (between 13 and 14.7 minutes), with ranges being 6–36 minutes, 6–38 minutes and 10–32 minutes for the K-S, A-D, and C-vM tests, respectively. The LSTM RNN returned an average TTR of 25.4 minutes. Notably, the LSTM produced the smallest range, with TTRs between 16.5 and 46 minutes (note the fractional minute is an artifact of averaging a given sample's TTR from multiple LSTM seeds) (**Figure S11**). Importantly, a statistically significant difference between these averages was only achieved between slope-based methods and the EDF methods, with similar methods (i.e. average slope *vs* single-point slope, or AD *vs* KS) generally displaying equivalent performance. We attribute this to the large variance in target titre between each patient sample. Considering the high sensitivities and average TTRs of 4.67, 4.67 and 6 minutes for K-S, A-D, and C-vM, respectively, it is clear that these methods prioritise rapid and accurate identification of positives. Despite the high confidence intervals (99.985%, 99.97%, and 99.99% for the K-S, A-D, and C-vM respectively) utilized for the three methods, TTRs were significantly shorter than the benchmark methods. These performance metrics can be regulated by adjusting test parameters, such as confidence interval, window length and consecutive run length. Although a higher confidence interval makes the test "stricter" in that the bar for a sample being identified a positive is higher, this may increase TTR and lower sensitivity. Conversely, a shorter run length or lower confidence interval may perform well on sensitivity, catching even low-titer (and thus low-signal) tests quickly, but at the cost of specificity, where higher signal negatives are misidentified as positive. In this work, the high confidence interval was selected to maximize the total number of samples correctly identified in the data set. This highlights the importance of considering the clinical implications of false positives or false negatives for the particular diagnostic assay and setting the relevant parameters accordingly (**Figure S12**).

Ranking the methods from lowest TTR (fastest) to highest TTR (slowest) for each sample (**Figure 5B**) supports the trends observed for the average TTRs. The EDF methods are consistently fastest, followed by the LSTM, then the slope-based methods, and finally the fluorescence magnitude method. In 20 out of 24 samples, the shortest (or joint-shortest) TTR was returned by an EDF test, with the Kolmogorov-Smirnov ranking 1st (or joint 1st) in 14 samples, the Anderson-Darling in 13, and the Cramèr-von Mises in ranking 1st in 6 samples. Interestingly, the four samples in which an EDF test did not provide the shortest TTRs are the ones in which no result was returned at all. In these cases, the LSTM was the only method that correctly identified these positives, suggesting superior sensitivity. At the other extreme, the fluorescence magnitude method yielded the shortest TTR in 8 out of the 10 samples it correctly identified (out of a total of 24), suggesting this analysis method is both the slowest and least sensitive.

**Figure 5C** provides a summary of the data obtained in this study and highlights the impact that different analysis methods have on performance metrics. By simply changing the analysis method, assay sensitivity ranged from 38% to 100%, specificity from 88% to 100%, and the average TTR from 4.7 minutes to 42.8 minutes. This strongly suggests that different classification approaches should be considered for different applications, depending on priorities. The Kolmogorov-Smirnov, Anderson-Darling, and Cramér-von Mises tests may be most appropriate if a highly sensitive and rapid assay is more important than assay specificity. For example, in screening of mild but infectious diseases, a test may optimize for rapid and accessible feedback, favouring false positives with a non-harmful intervention and the ability to retest over a false negative and disease spread. Conversely, the LSTM approach may be more appropriate if specificity is more important than sensitivity and assay time, and enough data is available to allow representative training of the network. The LSTM may be more relevant when diagnosing diseases such as cancer, where treatments can require the administration of potentially harmful therapies, a test should take a more conservative classification approach, minimizing false positives at all costs. Finally, our results highlight the impact of test parameters such as confidence interval, window length, and consecutive run length (number of consecutive calls necessary to trigger positive indication) on sensitivity, specificity, and time-to-result. Accordingly, these parameters should be carefully optimised for each individual assay.

# Conclusions

Whilst significant effort has focused on optimizing the chemistry and biology of CRISPR–Cas biosensing reactions, the analysis of CRISPR–Cas biosensing data has received relatively little attention despite its essential role within the IVD workflow. This work illustrates that statistical and ML methods have advantages in speed and accuracy over traditional fluorescence magnitude and slope-based approaches. Specifically, the K-S, A-D, and C-vM tests used in this work displayed optimal results across both performance metrics, having both the lowest TTR and the highest overall accuracy. The LSTM-RNN, whilst being slightly less sensitive, displayed superior specificity than the statistical methods we tested, and outperformed traditional slope-based methods. Whilst we acknowledge the importance of increasing the relatively small data training datasets used in this work, we have shown that RNNs can be used as powerful classifiers in CRISPR–Cas diagnostics and that future application to large scale datasets has significant potential.

It is important to stress that we take no stance on the "best" optimization strategy. Rather, we simply aim to establish the idea that that the choice of analysis method is equally as important as the inherent CRISPR–Cas biosensor performance. Considering the rapid growth of ML methods, we hope that this work sparks interest in using other ML methods for the analysis and classification of CRISPR–Cas (and other) biosensing data. One potential avenue in this regard would be to explore different approaches for different aspects of assay classification (i.e. one method for classifying positive *vs* negative, and another for quantification of analyte concentration), and then employing these in parallel. Further avenues involve employing a mixture of expert architecture, a technique recently employed in the field of large language models to much success[54,55]. In the context of CRISPR–Cas biosensors, this could involve employing a mixture where some models ("experts") excel in rapidly calling high-signal positives and others in differentiating low-signal positives *vs* negatives. At each time point a router network would select the experts most suited to analyze the input data. Regardless of the route, the data presented herein suggests that applying concepts from machine learning to CRISPR–Cas diagnostics represents a promising and direct route for improving performance. It is our hope that others adopt and adapt these methods in their own work to broadly improve the performance of CRISPR-Cas-based diagnostic assays.

## Supporting Information

Additional details of the statistical and LSTM methods, including optimizations of window length, run length, and confidence intervals, and analyses of the individual seeds of the LSTM.

## Acknowledgments

The authors would like to acknowledge that **Figure 1** was created with BioRender.com. DAR acknowledges funding from the ETH Career Seed Grant (SEED-13 21-2).

## Contributions

Conceptualisation: JML, NK; Data analysis JML, NK, MH; Writing (original draft): JML, NK; Writing (review and editing): DAR, JML, NK, MH, AJdM; Data visualisation: JML, NK, MH, DAR; Supervision: DAR, AJdM; Resources: AJdM; Funding acquisition: AJdM.

# Machine Learning and statistical classification of CRISPR-Cas12a diagnostic assays

Nathan Khosla[1][#], Jake M. Lesinski[1][#], Marcus Haywood-Alexander[2], Andrew J. deMello[1*] & Daniel A. Richards[1*]

[1]Institute for Chemical and Bioengineering, ETH Zurich, Vladimir-Prelog-Weg 1, 8093, Zürich, Switzerland.

[2]Institute of Structural Engineering, ETH Zurich, Stefano-Franciscini-Platz 5, 8049, Zürich, Switzerland.

[#] These authors contributed equally

* daniel.richards@chem.ethz.ch, andrew.demello@chem.ethz.ch,

# Electronic Supporting Information

# Reference Standard Methods

**Table S1.** Times-to-result obtained using fluorescence magnitude, average slope, and single-point slope analytical methods. Anyplex results were sourced from Lesinski et al.[1]

|  | PCR | Time to Result (min) | | | | | | | | |
|---|---|---|---|---|---|---|---|---|---|---|
|  | Anyplex | Endpoint | | | Average Slope | | | Single-Point Slope | | |
|  | *Qualitative* | *Trial 1* | *Trial 2* | *Trial 3* | *Trial 1* | *Trial 2* | *Trial 3* | *Trial 1* | *Trial 2* | *Trial 3* |
| Positive 1 | +++ | 32 | 30 | 30 | 28 | 26 | 26 | 26 | 24 | 24 |
| Positive 2 | ++ | 58 | 80 | FN | 36 | 42 | FN | 54 | FN | FN |
| Positive 3 | ++ | FN | FN | FN | 42 | FN | FN | FN | FN | FN |
| Positive 4 | ++ | FN | FN | FN | 46 | 54 | FN | FN | FN | FN |
| Positive 5 | ++ | FN | FN | FN | 40 | 90 | FN | FN | FN | FN |
| Positive 6 | ++ | 32 | 36 | FN | 28 | 30 | FN | 28 | 30 | FN |
| Positive 7 | +++ | FN | 32 | 30 | 84 | 28 | 26 | FN | 26 | 26 |
| Positive 8 | ++ | FN | FN | 68 | 34 | FN | 42 | FN | FN | 68 |

FN = False Negative

# Statistical Methods

The three statistical methods utilize empirical cumulative distribution functions (ECDFs) originating from pre-established populations comprised of known sets of negatives. In each case, the ECDF of the sample of interest is compared to the ECDF of the negatives and categorized accordingly.

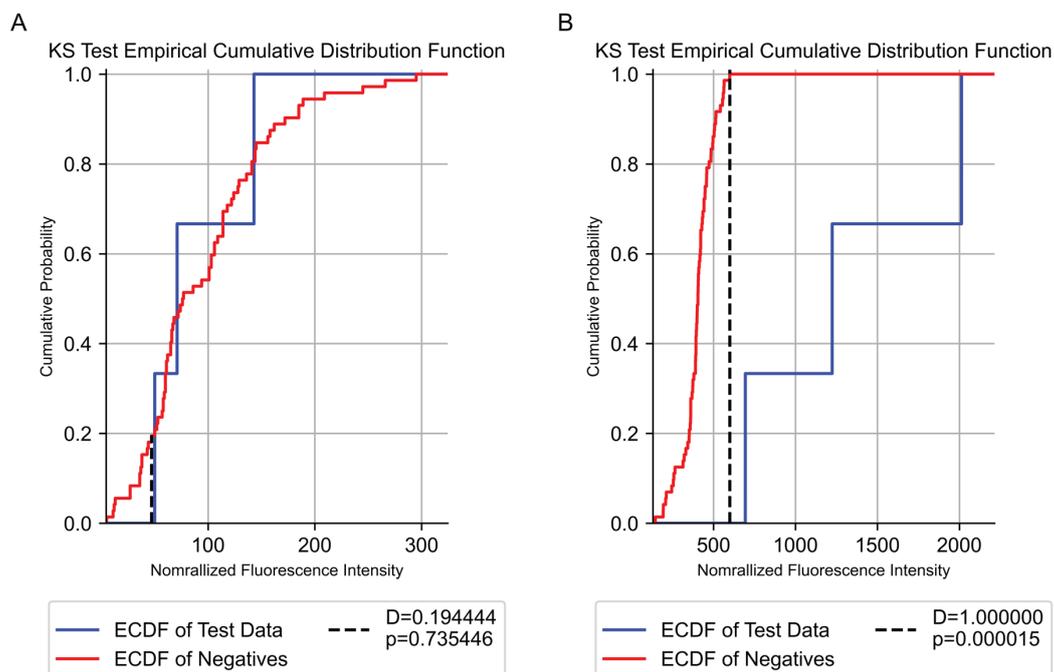

**Figure S1**. Visualization of the underlying functions of the K-S test. The ECDF of the larger negative data population (red line) is compared to the ECDF of the test of interest (blue line) at each time point. A) compares the two ECDFs are the start of the test while B) illustrates greater separation of the two ECDFs at the time when the test data is classified.

The statistical methods employed in this paper, namely the Kolmogorov-Smirnov (K-S), Anderson-Darling (A-D), and Cramer-von Mises (C-vM) test, were evaluated against given positive/negative calls from PCR testing from Lesinski et al, as seen below (**Table S2**).

**Table S2.** Times-to-result obtained using the Kolmogorov-Smirnov, Anderson-Darling, and Cramér von-Mises tests. Anyplex results were sourced from Lesinski et al.[1]

| | PCR | Time to Result (min) | | | | | | | | |
| --- | --- | --- | --- | --- | --- | --- | --- | --- | --- | --- |
| | Anyplex | Kolmogorov-Smirnov | | | Anderson-Darling | | | Cramér-von Mises | | |
| | *Qualitative* | *Trial 1* | *Trial 2* | *Trial 3* | *Trial 1* | *Trial 2* | *Trial 3* | *Trial 1* | *Trial 2* | *Trial 3* |
| Positive 1 | +++ | 26 | 24 | 24 | 26 | 24 | 24 | 32 | 30 | 30 |
| Positive 2 | ++ | 6 | 6 | FN | 6 | 6 | FN | 10 | 10 | FN |
| Positive 3 | ++ | 6 | 6 | 6 | 6 | 6 | 6 | 10 | 10 | 10 |
| Positive 4 | ++ | 6 | 8 | FN | 6 | 8 | FN | 10 | 10 | FN |
| Positive 5 | ++ | 6 | FN | 8 | 6 | FN | 8 | 10 | 10 | 10 |
| Positive 6 | ++ | 8 | 26 | FN | 6 | 26 | FN | 10 | 10 | FN |
| Positive 7 | +++ | 10 | 24 | 24 | FN | 24 | 24 | 10 | 10 | 10 |
| Positive 8 | ++ | 6 | 8 | 36 | 6 | FN | 38 | 10 | 10 | 10 |

FN = False Negative

Each statistical method involved three adjustable parameters, namely run length, window length, and alpha. The run length determines the number of consecutive positive categorizations that must occur in a row for the sample to be considered positive. The window length defines the sliding range of time points considered in the test evaluation. At each instance in time a stretch of data is considered. Alpha represents the significance threshold. A sweep of these three parameters was performed for each statistical method, and the total accuracy, sensitivity, and specificity was evaluated for each parameter set (**Figures S2-10**).

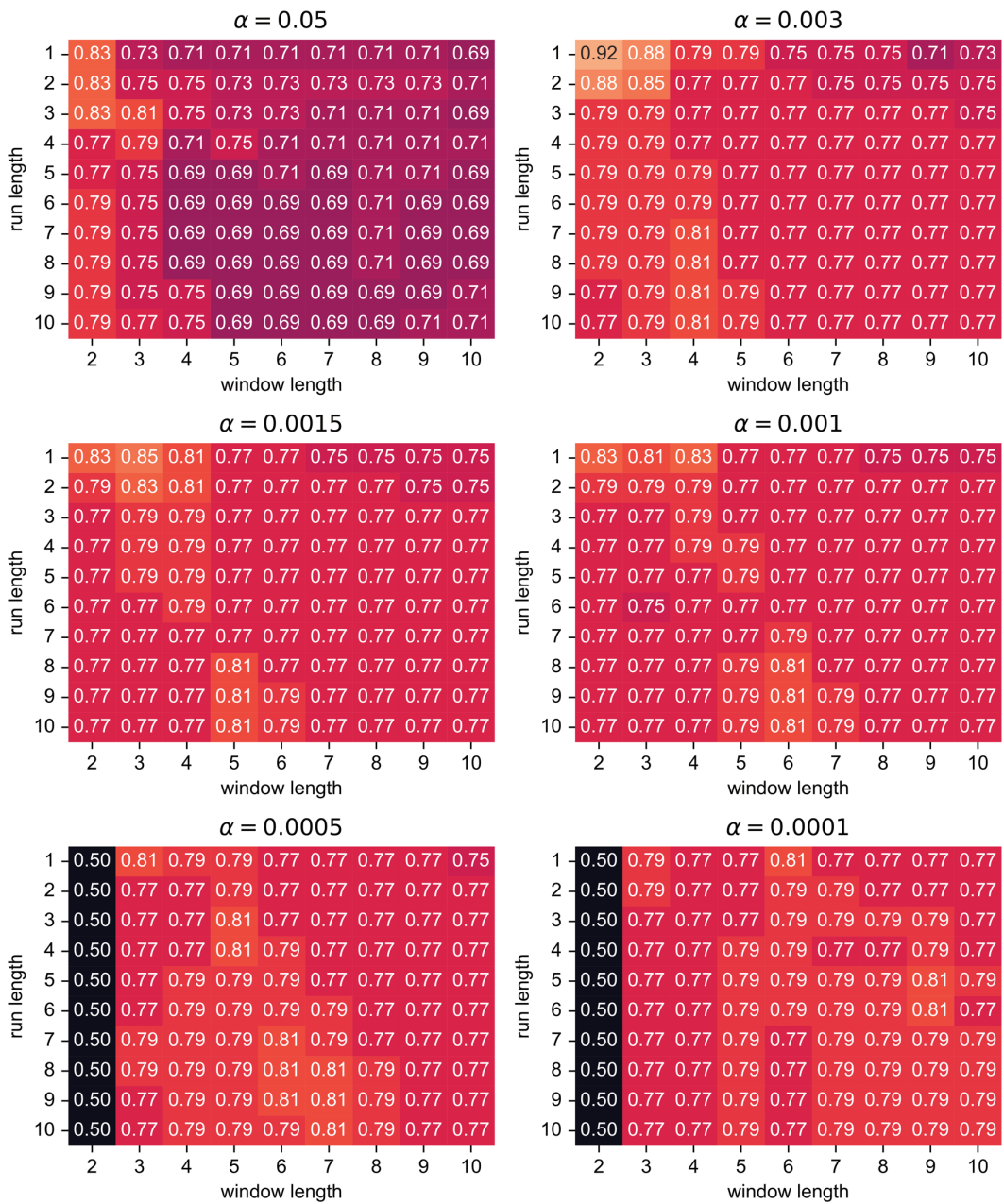

**Figure S2.** Variation of the underlying parameters used in the K-S test. The number in each cell indicates total test accuracy.

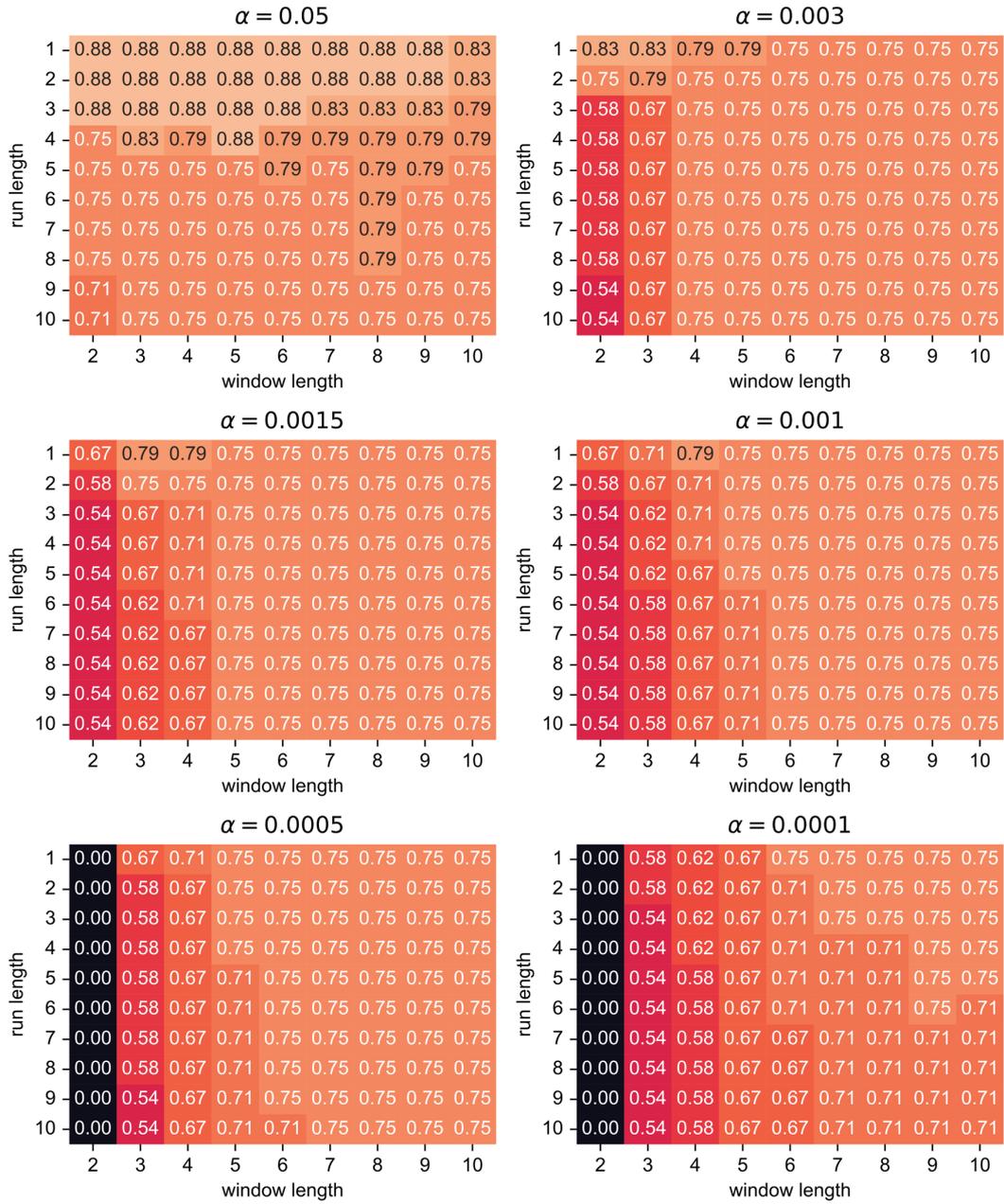

**Figure S3.** Variation of the underlying parameters used in the K-S test. The number in each cell indicates sensitivity.

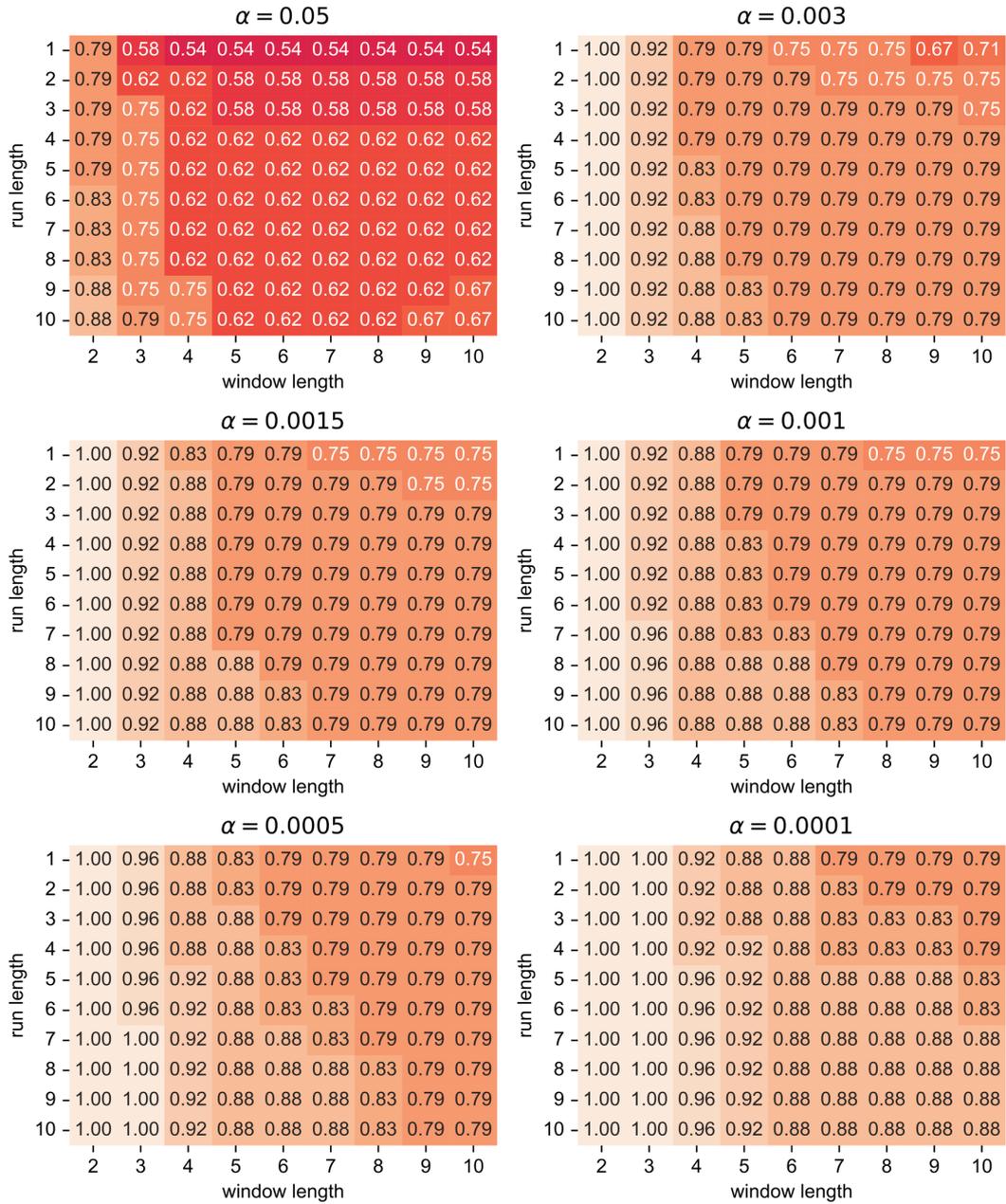

**Figure S4.** Variation of the underlying parameters used in the K-S test. The number in each cell indicates specificity.

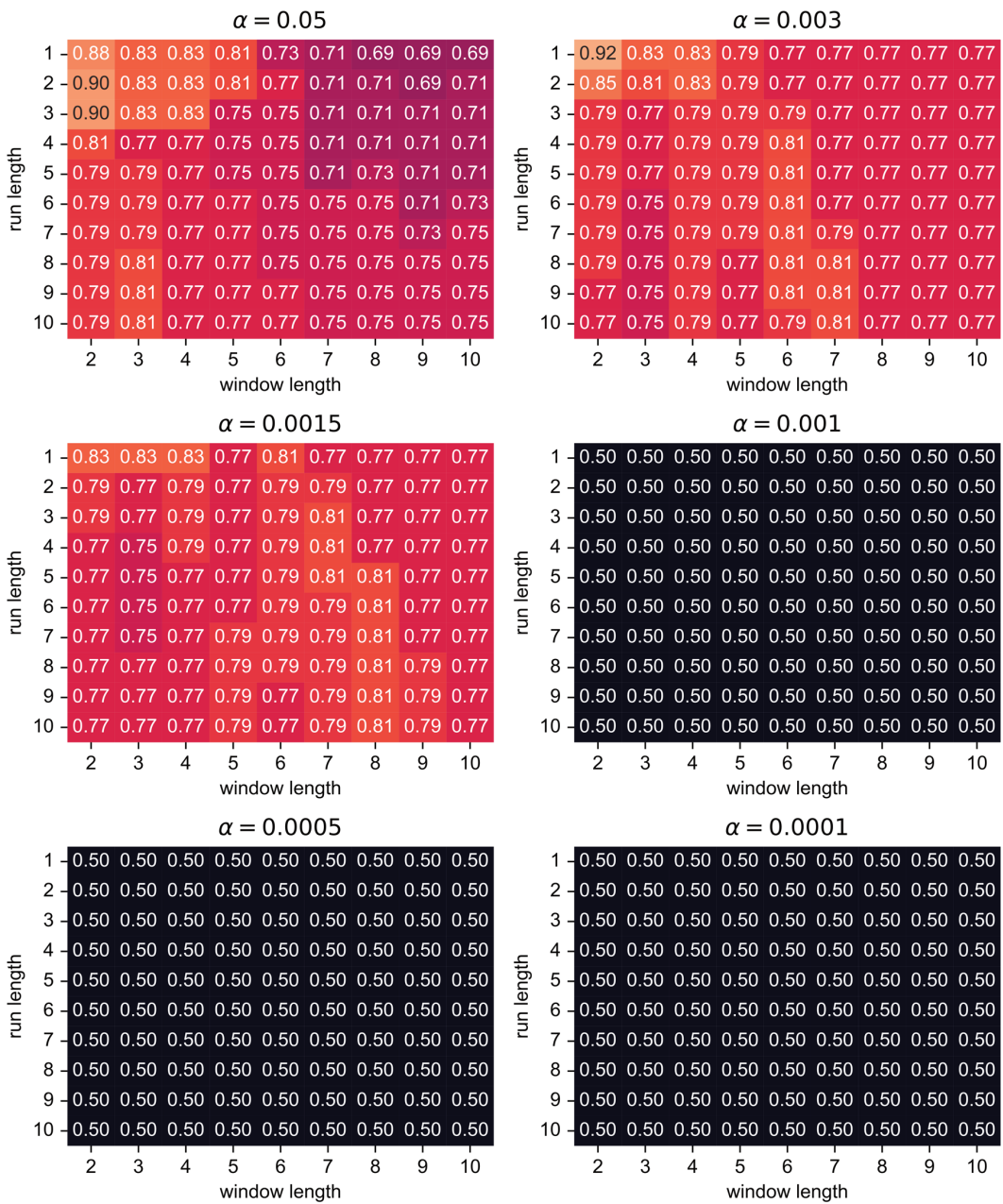

**Figure S5.** Variation of the underlying parameters used in the A-D test. The number in each cell indicates total test accuracy.

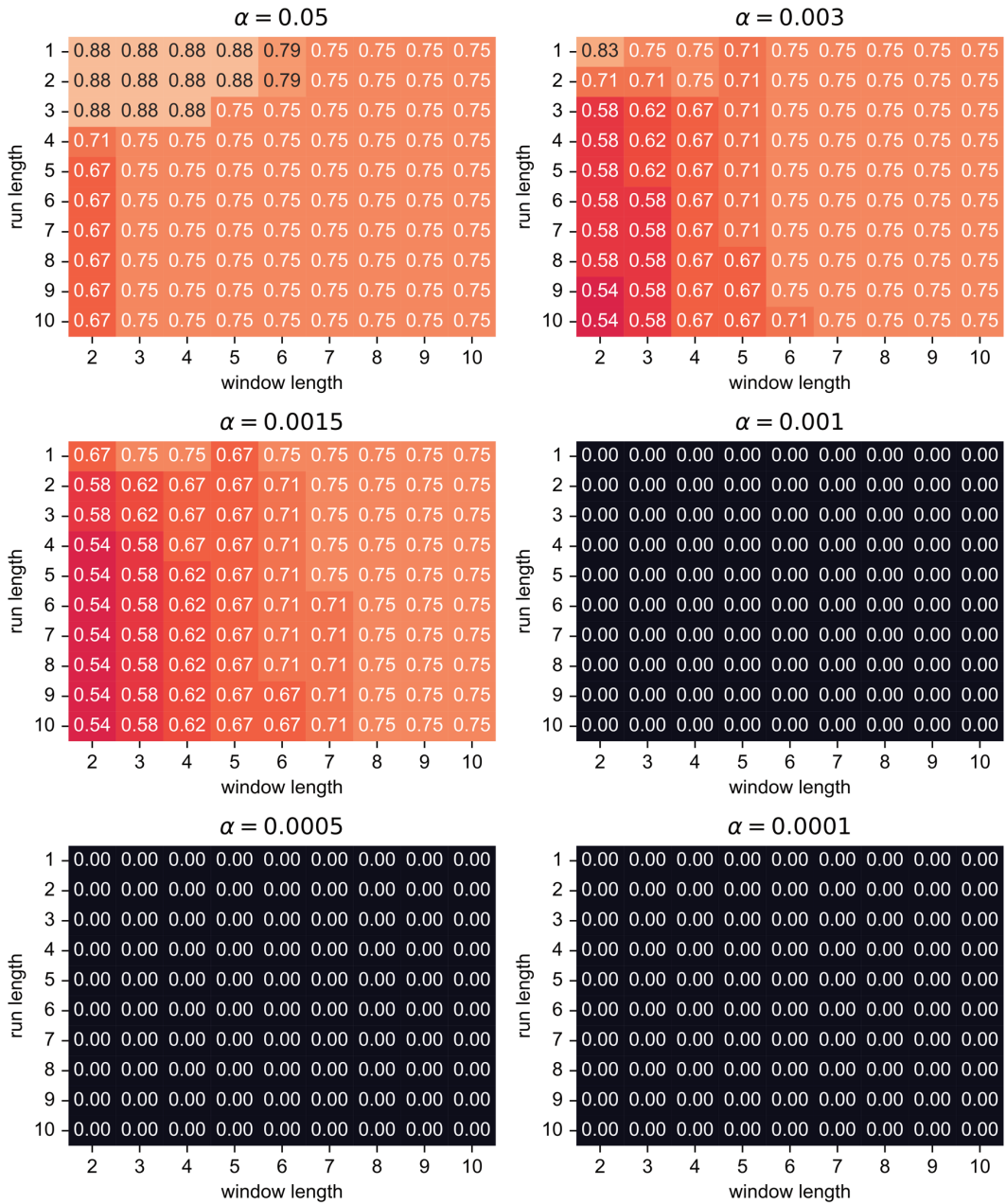

**Figure S6.** Variation of the underlying parameters used in the A-D test. The number in each cell indicates sensitivity.

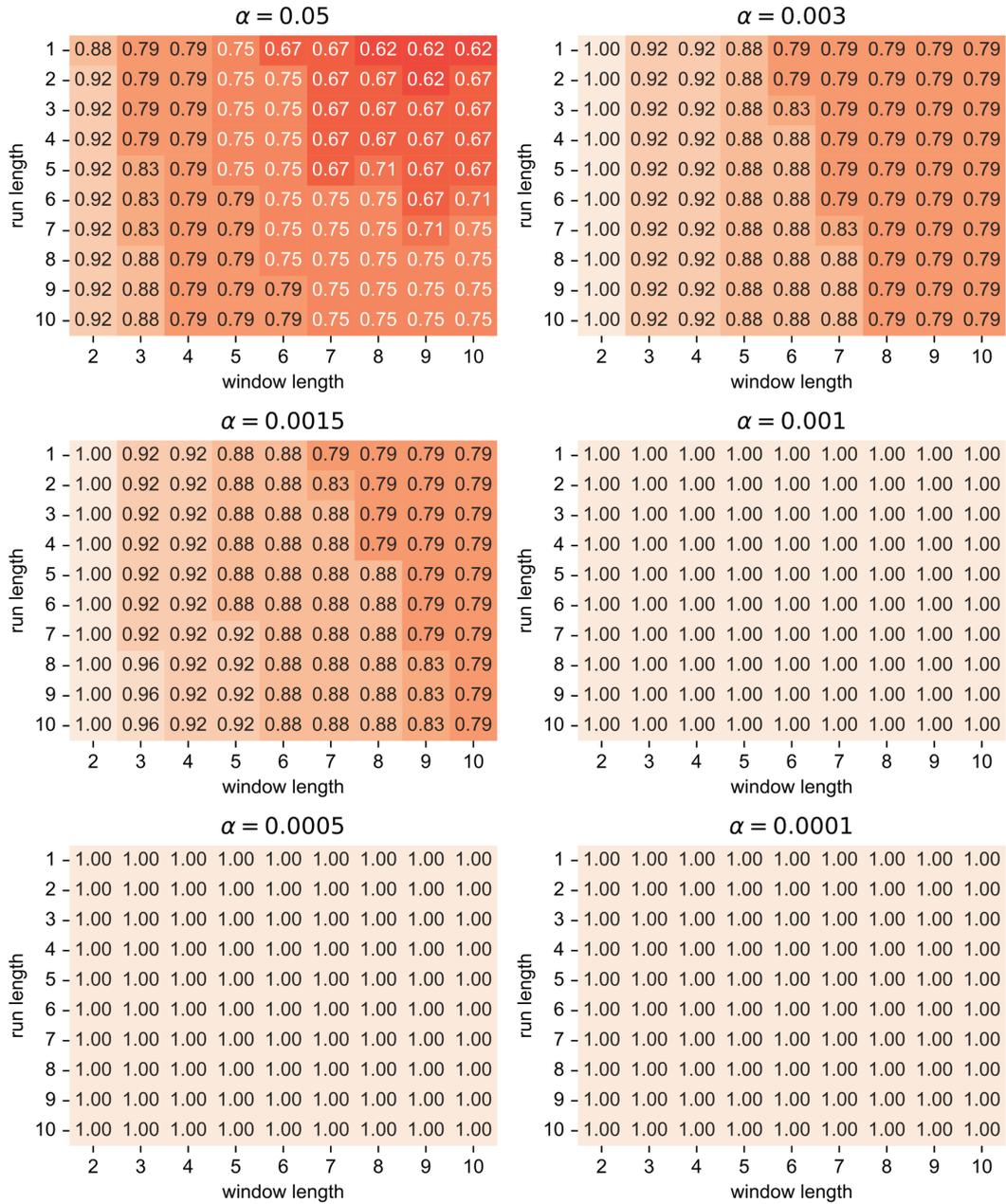

**Figure S7.** Variation of the underlying parameters used in the A-D test. The number in each cell indicates specificity.

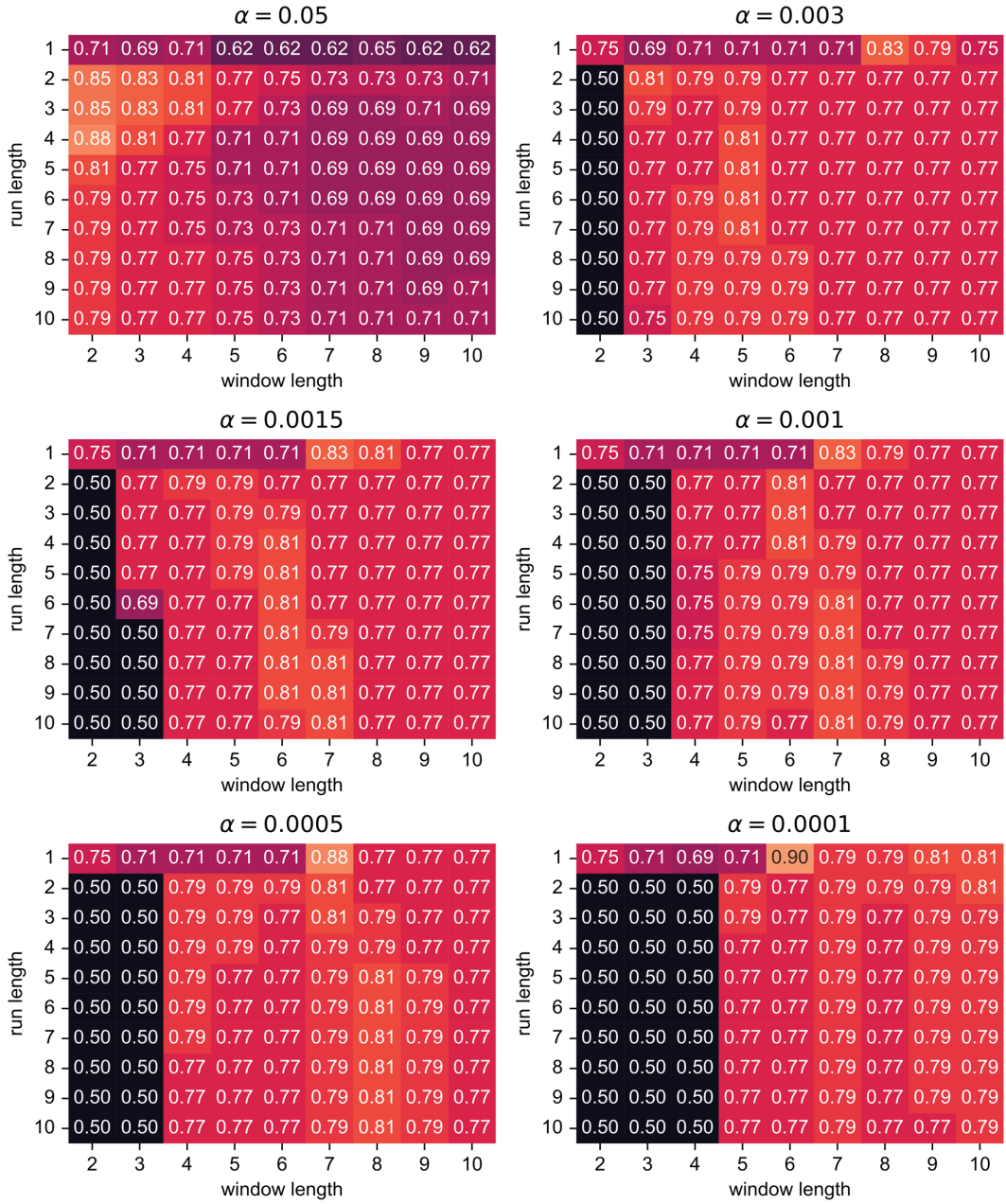

**Figure S8.** Variation of the underlying parameters of the C-vM test. The number in each cell indicates total test accuracy.

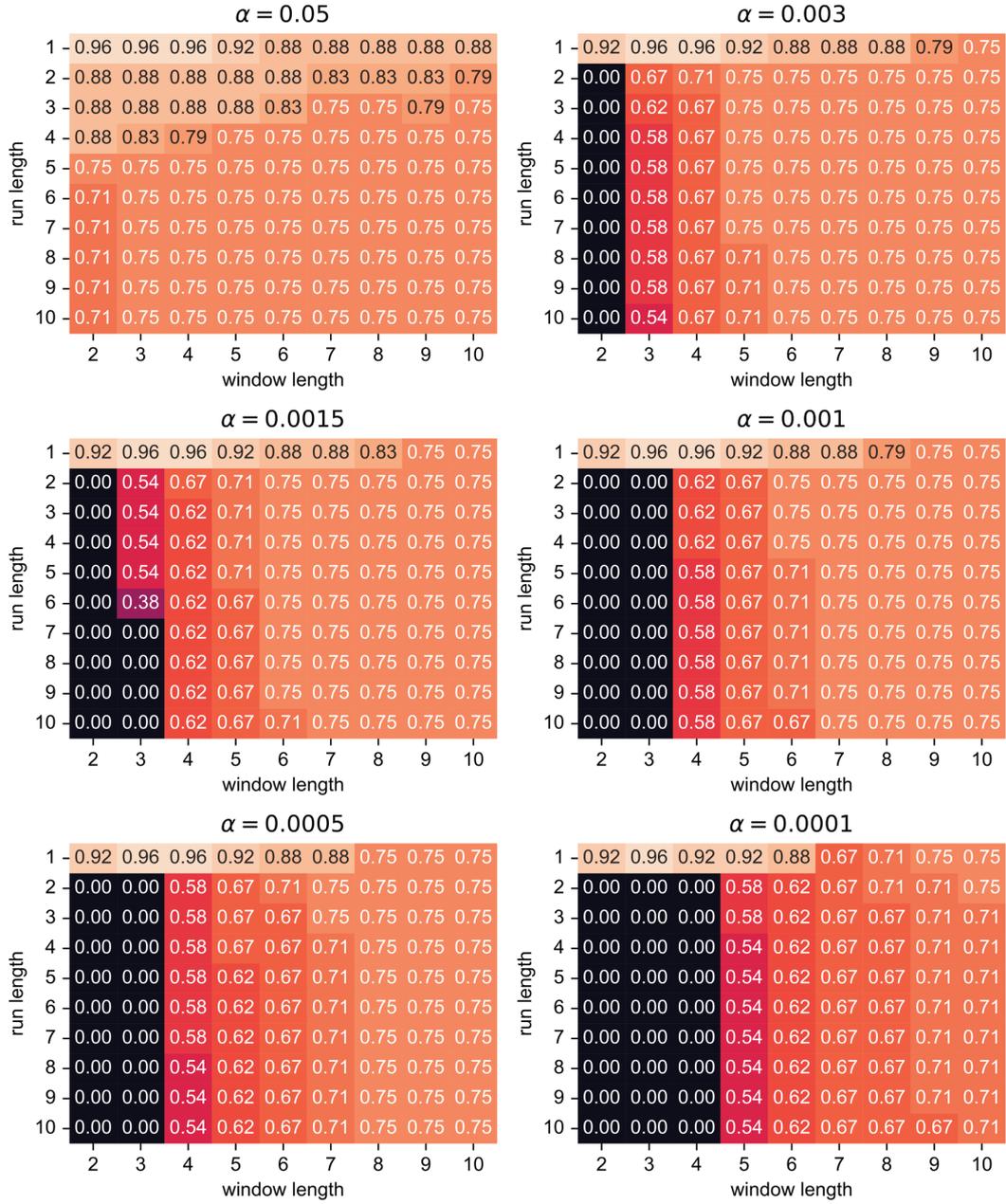

**Figure S9.** Variation of the underlying parameters of the C-vM test. The number in each cell indicates test sensitivity.

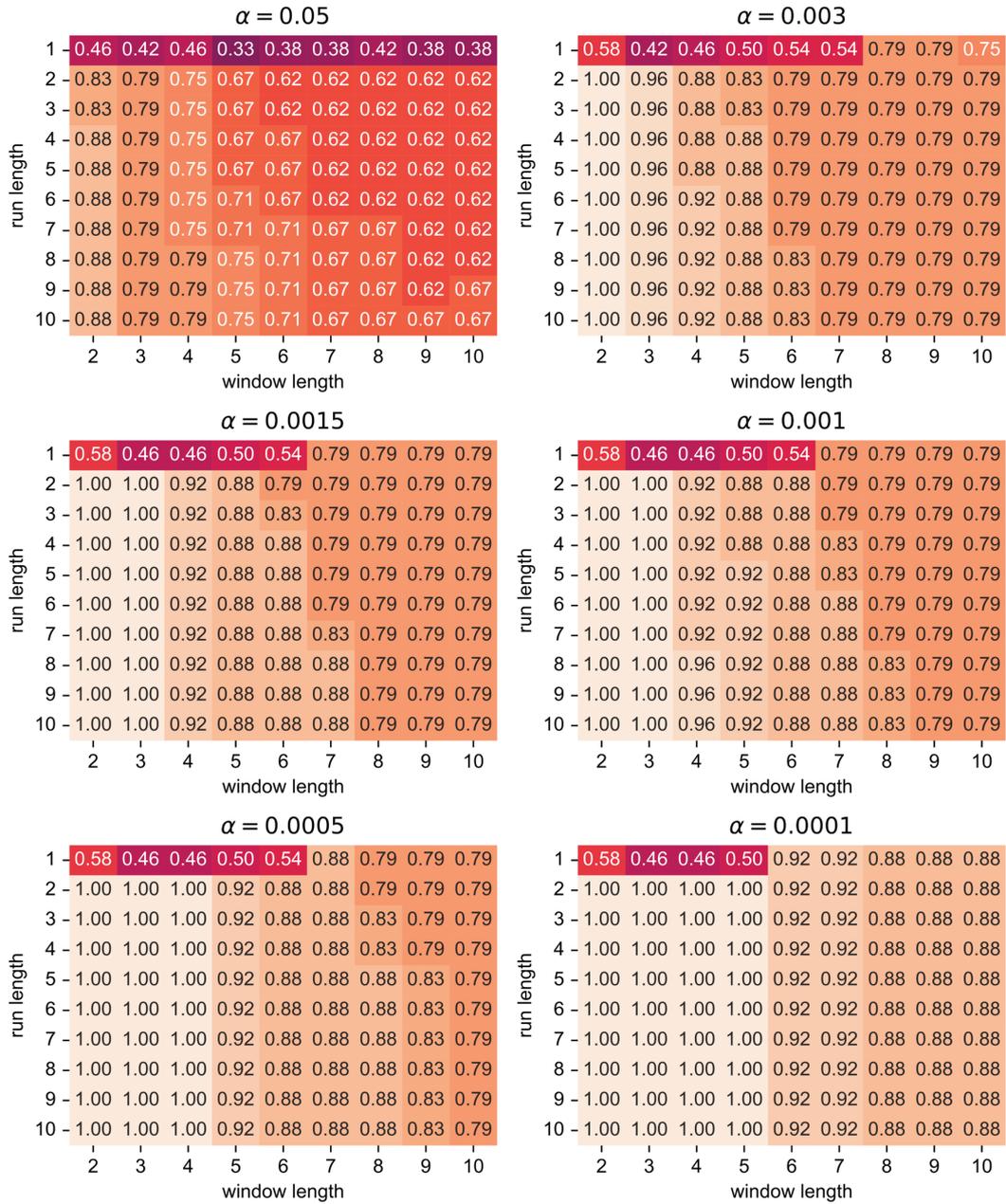

**Figure S10.** Variation of the underlying parameters of the C-vM test. The number in each cell indicates test specificity.

# Long Short-Term Memory (LSTM) network

**Table S3.** The different seeds utilized in the LSTM. Multiple seeds were employed to reduce the effects of random chance.

|         | Seed Value |
|---------|------------|
| Seed 1  | 424242     |
| Seed 2  | 93407      |
| Seed 3  | 92646      |
| Seed 4  | 92037      |
| Seed 5  | 43810      |
| Seed 6  | 23777      |
| Seed 7  | 15228      |
| Seed 8  | 8009       |
| Seed 9  | 123        |
| Seed 10 | 314159     |

The LSTM runs a prediction on a sample at a given time and computes a confidence value that the sample is positive. While naïve classifiers may identify a positive if the model is >0.5 confident that the sample is positive, a more conservative cutoff was used in the current study. Cutoffs of 0.90, 0.95, and 0.99 were tested, with the one yielding the highest total accuracy being selected (**Figure S11**).

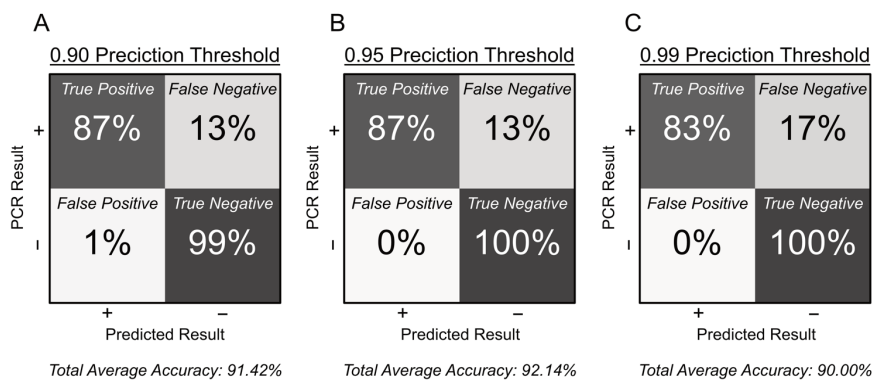

**Figure S11.** Confusion matrices illustrating how sensitivity and specificity are impacted by variation in prediction threshold used within the LSTM model.

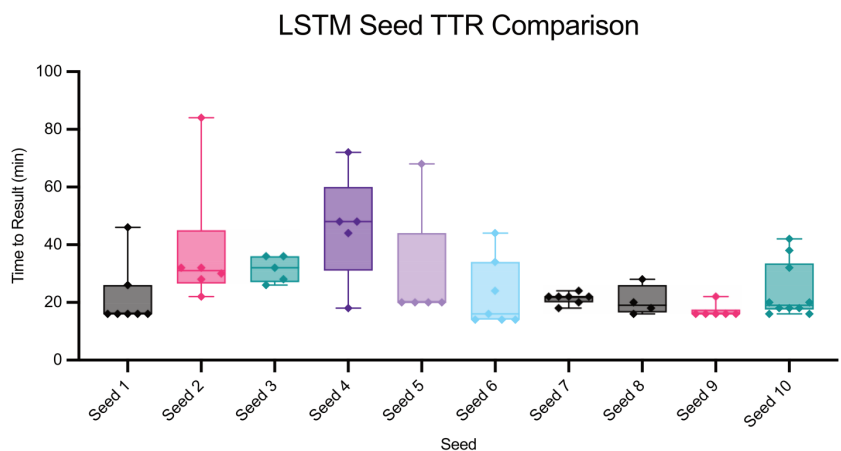

**Figure S12**. Comparison of TTR between LSTM seeds. Error bars represent the 95% CI.